\documentclass[article,nofootinbib,notitlepage]{revtex4-1}
\usepackage{amssymb}
\usepackage{amsfonts}
\usepackage{amsmath}
\usepackage{graphicx}
\usepackage{dcolumn}
\usepackage{bm}
\usepackage[latin1]{inputenc}
\usepackage[dvips]{hyperref}

\begin{document}

\title{Approximated solutions to Born-Infeld dynamics}
\author{Rafael Ferraro\bigskip}
\email{ferraro@iafe.uba.ar}
\thanks{member of Carrera del Investigador Cient\'{\i}fico (CONICET, Argentina). }
\affiliation{Instituto de Astronom\'\i a y F\'\i sica del Espacio
(IAFE, CONICET-UBA), Casilla de Correo 67, Sucursal 28, 1428 Buenos
Aires, Argentina.} \affiliation{Departamento de F\'\i sica, Facultad
de Ciencias Exactas y Naturales, Universidad de Buenos Aires, Ciudad
Universitaria, Pabell\'on I, 1428 Buenos Aires, Argentina.}
\affiliation{Departamento de F\'\i sica, Facultad de Ciencias
Exactas y Naturales, Universidad de Buenos Aires, Ciudad
Universitaria, Pabell\'on I, 1428 Buenos Aires, Argentina.}

\author{Mauro Nigro\bigskip} \email{nigro@df.uba.ar}
\affiliation{Departamento de F\'\i sica, Facultad de Ciencias
Exactas y Naturales, Universidad de Buenos Aires, Ciudad
Universitaria, Pabell\'on I, 1428 Buenos Aires, Argentina.\bigskip}

\begin{abstract}
The Born-Infeld equation in the plane is usefully captured in
complex language. The general exact solution can be written as a
combination of holomorphic and anti-holomorphic functions. However,
this solution only expresses the potential in an implicit way. We
rework the formulation to obtain the complex potential in an
explicit way, by means of a perturbative procedure. We take care of
the secular behavior common to this kind of approach, by resorting
to a symmetry the equation has at the considered order of
approximation. We apply the method to build approximated solutions
to Born-Infeld electrodynamics. We solve for BI electromagnetic
waves traveling in opposite directions. We study the propagation at
interfaces, with the aim of searching for effects susceptible to
experimental detection. In particular, we show that a reflected wave
is produced when a wave is incident on a semi-space containing a
magnetostatic field.
\end{abstract}

\maketitle

\section{Introduction}

\label{sec:intro}

Born-Infeld (BI) electrodynamics \cite{Born1} is a non-linear generalization
of Maxwell's theory. BI action governs the tensor field $F_{ij}=\partial
_{i}A_{j}-\partial _{j}A_{i}$; it is defined as
\begin{equation}
\mathcal{S}[A_{i}]\ =\ \frac{b^{2}}{4\,\pi }\ \int ~d^{4}x~\left( 1-\sqrt{1+%
\frac{2S}{b^{2}}-\frac{P^{2}}{b^{4}}}\right) \ ,  \label{action}
\end{equation}%
where $S$ and $P$ are the scalar and pseudo-scalar
\begin{equation}
S\ \equiv \ \frac{1}{4}\ F_{ij}\ F^{ij}\ =\ \frac{1}{2}\ (B^{2}-E^{2})~,%
\hspace{0.5in}P\ \equiv \ \frac{1}{4}\ \ast \hfill \!{}F_{ij}\ F^{ij}\ =\
\mathbf{E\cdot B~}.  \label{SP}
\end{equation}%
BI dynamical equations in vacuum are%
\begin{equation}
\partial _{i}\mathcal{F}^{ij}\ =\ 0\ ,\hspace{0.5in}\mathrm{where}\hspace{%
0.5in}\mathcal{F}_{ij}\ \equiv \ \frac{F_{ij}-\frac{P}{b^{2}}\ \ast \hfill
\!{}F_{ij}}{\sqrt{1+\frac{2S}{b^{2}}-\frac{P^{2}}{b^{4}}}}\ .  \label{BIdyn}
\end{equation}%
Constant $b$ in Eqs.~(\ref{action}) and (\ref{BIdyn}) has units of field; it
represents the scale for which BI theory departs from Maxwell's theory. In
the weak field regime both theories lead to similar results, since $\mathcal{%
F}^{ij}$ in Eq.~(\ref{BIdyn}) goes to ${F}^{ij}$ in the limit $b\rightarrow
\infty $ (see also References \cite{Born2,Born3,Born4}).

\bigskip

Although the historical motivation for BI theory was to circumvent
the infinite self-energy of Maxwell's point-like charge (by
enforcing the electric field to have the value $b$ at the position
of the charge), nowadays the interest in this kind of theory has
moved towards high-energy physics, since the low energy limit of
string theories contains BI-like Lagrangians
\cite{Fradkin,Abou,Leigh,Met,Tse}. Except for a case without
physical interest, BI theory is the only non-linear electrodynamics
free of
birefringence and shock waves, and displaying causal propagation \cite%
{Plebanski,Boillat,Deser,Bial,Kerner}. Similarities between BI dynamics and
MHD equations have been considered in Ref. \cite{Brenier}.

\bigskip

Apart from the trivial spherically symmetric case, exact stationary BI
configurations are hard to obtain; they have been studied in the form of
perturbative series \cite{Kies1,Kies2}. However, BI electrostatics in the
plane can be exactly solved, although in an implicit way. After replacing $%
\mathbf{E} =-\nabla u(x,y)$ and $\mathbf{B}=0$ in Eq.~(\ref{BIdyn}), one
obtains the equation governing the electrostatic potential $u(x,y)$:
\begin{equation}
(1\,-\,b^{-2}\,u_{,y}^{2})\ u_{,xx}+2\,b^{-2}\,u_{,y}\ u_{,x}\
u_{,yx}+(1\,-\,b^{-2}\,u_{,x}^{2})\ u_{,yy}\ =\ 0\ ,  \label{BI}
\end{equation}%
Equation (\ref{BI}) can be regarded as a non-linear extension of the
Laplace equation $\Delta u(x,y)=0$. While the Laplace equation in
the plane immediately refers to holomorphic functions, since the
real and imaginary parts of any holomorphic function verify Laplace
equation, the relation between the solutions to Eq.~(\ref{BI}) and
holomorphic functions is much more subtle. This relation was
obtained by Pryce \cite{Pryce, Pryce1} (see also References
\cite{Ferraro, Ferraro0, Ferraro1, Ferraro2}). It is worth
mention that Eq.~(\ref{BI}) can be also obtained directly from the action%
\begin{equation}
\mathcal{S}[u]\ =\ \frac{b^{2}}{4\,\pi }\ \int ~dx\wedge dy~\sqrt{%
1-b^{-2}\left( u_{,x}^{\ 2}+u_{,y}^{\ 2}\right) }\ .  \label{actionu}
\end{equation}%
This action coincides with the one for the problem of \textit{maximal
surfaces }in (2+1) Minkowski space (space-like surfaces with vanishing mean
curvature). It is well known that the solution to this geometrical problem
can be written in a parametric way; if $t=b^{-1}\,u(x,y)$ is the equation
for the surface defined on a domain $D$ of the complex plane $z=x+i~y$, then
the solutions admit the Weierstrass-Enneper parametrization \cite{Kobayashi}:%
\begin{equation}
\left( x(\varsigma ),y(\varsigma ),\ t(\varsigma )\right) =\text{Re}\ \int
\left( \frac{1}{2}\,\phi \,(1+\varphi ^{2}),\ \frac{i}{2}\,\phi \,(1-\varphi
^{2}),\ -\phi \,\varphi \right) \ d\varsigma \ ,  \label{WE}
\end{equation}%
where $\phi $ is holomorphic and $\varphi $ is meromorphic on $D$ such that $%
\phi \,\varphi ^{2}$ is holomorphic on $D$ and $|\varphi (z)|\neq 1$ for $%
z\in D$. Like the Weierstrass-Enneper parametrization, also the method of
References \cite{Pryce, Pryce1,Ferraro, Ferraro0, Ferraro1,Ferraro2} is
unable of offering the solutions in an explicit way. Instead, the solution $%
u(x,y)=u(z, \overline{z})$ is given in an implicit way through a function $%
z=z(w, \overline{w})$, $u$ being the real part of $w$.

\bigskip

Equation (\ref{BI}) becomes a wave equation by replacing $y$ with
$i\,t$. At the level of the action (\ref{actionu}), this procedure
leads to a descendent of the Nambu-Goto action in 2+1 dimensions
\cite{Bordemann}, which straightforwardly connects BI wave equation
with string theory. Like its static relative (\ref{BI}), the
Born-Infeld wave equation for the scalar potential $u(t,x)$ also
provides solutions to some configurations of BI electrodynamics. We
remark that, as a consequence of the non-linearity, BI
electromagnetic waves interact with other waves and static fields \cite%
{Barbashov,Salazar,Novello,Ferraro3,Aiello,Ferraro4}. Only free plane waves
evolve like in Maxwell's theory, because $S$ and $P$ vanish in such case
(compare with the cylindrical wave in Ref. \cite{Petrov}).

\bigskip

Due to its relevance both for static and propagating BI electromagnetic
configurations, together with the impact in related theories, the methods to
get solutions to Eq.~(\ref{BI}) are deserving of our attention. The rest of
the paper is organized as follows. In Section \ref{sec2} we show that action
(\ref{actionu}) can be transformed into a simple form quadratic in the
derivatives, when written in terms of the fields $z=z(w, \overline{w})$. In
Section \ref{sec3} we invert the exact linear equation accomplished by $%
z=z(w, \overline{w})$ to get a first order differential equation for
the complex potential $w=w(z,\overline{z})$, which is valid at the
first order in $b^{-2}$. In Section \ref{sec4} we solve for periodic
$w(z,\overline{z})$ functions, and heal the unbounded behavior
characterizing this kind of perturbative approach. The healing
method takes advantage of a symmetry of the
differential equation at the considered order of approximation. In Section %
\ref{sec5} we solve the BI scalar wave equation for a configuration
of two waves traveling in opposite directions. The interaction
between these waves reveals as a modulation of the amplitudes and a
shift in the dispersion relation. As shown in Section \ref{sec6},
this solution can be exploited to obtain configurations of equally
polarized BI electromagnetic waves traveling in opposite directions;
in particular we get the expression for a BI electromagnetic wave
interacting with a mirror. We also show that a BI electromagnetic
wave entering a region containing a uniform magnetic field is
reflected at the interface and changes its polarization; this is an
effect typical of a non-linear theory, susceptible of experimental
detection. In Section \ref{sec7} we display the conclusions.

\section{Complex formulation}

\label{sec2} We will change from Cartesian coordinates to complex
coordinates in the plane: $(x,y)\longrightarrow (z,\overline{z})$.
So, it follows that
\begin{equation}
dx^{2}+dy^{2}=|dz|^{2}=dz~d\overline{z}~,
\end{equation}%
which means that the metric tensor in the chart $(z,\overline{z})$ has the
components%
\begin{equation}
g_{zz}=0=g_{\overline{z}\overline{z}}~,\hspace{0.25in}g_{z\overline{z}}=%
\frac{1}{2}=g_{\overline{z}z}~.
\end{equation}%
By replacing $u_{,x}^{\ 2}+u_{,y}^{\ 2}=g^{\mu \nu }\,u_{,\mu }\,u_{,\nu
}=4\,u_{z}\,u_{\overline{z}}$ (the subindexes indicate partial derivatives)
the action (\ref{actionu}) becomes%
\begin{equation}
\mathcal{S}[u]\ =\ \frac{b^{2}}{8\,\pi \,i}\ \int ~dz\wedge d\overline{z}~%
\sqrt{1-4b^{-2}\ u_{z}\,u_{\overline{z}}}\ .  \label{actionuz}
\end{equation}%
From this action we can reobtain the equation (\ref{BI}) written in terms of
the derivatives $u_{z}$, $u_{\overline{z}}\ $:
\begin{equation}
\ u_{z\overline{z}}\ +\frac{1}{\,b^{2}}\ \,u_{\overline{z}}^{2}\ \ u_{zz}-%
\frac{2}{\,b^{2}}\ u_{z}\ \ u_{\overline{z}}\ \ u_{z\overline{z}}+\frac{1}{%
\,b^{2}}\ \,u_{z}^{2}\ \ u_{\overline{z}\overline{z}}=\ 0~.  \label{BIz}
\end{equation}%
The equipotential lines $u(x,y)=\mathit{constant}$ can be regarded as
coordinate lines of a non-Cartesian chart; so we will introduce a change of
coordinates $(z,\overline{z})\longrightarrow (w,\overline{w})$, where the
real and imaginary parts of $w=u+i~v$ are coordinates $(u,v)$ in $%
\mathbb{R}
^{2}$. The Jacobian matrix is%
\begin{equation}
\left(
\begin{array}{cc}
u_{z} & u_{\overline{z}} \\
v_{z} & v_{\overline{z}}%
\end{array}%
\right) ~=~\left(
\begin{array}{cc}
\frac{\partial z}{\partial u} & \ \ \ \frac{\partial z}{\partial v} \\
&  \\
\frac{\partial \overline{z}}{\partial u} & \ \ \ \frac{\partial \overline{z}%
}{\partial v}%
\end{array}%
\right) ^{-1}~=~\left(
\begin{array}{cc}
(z_{w}+z_{\overline{w}}) & \ \ \ \ \ i\left( z_{w}-z_{\overline{w}}\right)
\\
&  \\
(\overline{z}_{w}+\overline{z}_{\overline{w}}) & \ \ \ \ \ i\left( \overline{%
z}_{w}-\overline{z}_{\overline{w}}\right)%
\end{array}%
\right) ^{-1}~.  \label{matrix}
\end{equation}%
This coordinate change is not unique, since we are free to choose
the coordinate $v$. We are going to show that BI equation
(\ref{BIz}) is remarkably simple when written not for
$u(z,\overline{z})$ but for $z(w, \overline{w})$, provided a
convenient coordinate $v$ is chosen. Let us rewrite the pieces
entering the action (\ref{actionuz}). On the one hand it is
\begin{equation}
dz\wedge d\overline{z}~=~\left( z_{w}~\overline{z}_{\overline{w}}-z_{%
\overline{w}}~\overline{z}_{w}\right) ~dw\wedge d\overline{w}~.
\end{equation}%
On the other hand we can solve $u_{z}$ from Eq.~(\ref{matrix}):%
\begin{equation}
u_{z}~=\frac{1}{2}~\frac{\overline{z}_{w}-\overline{z}_{\overline{w}}}{z_{%
\overline{w}}~\overline{z}_{w}-z_{w}~\overline{z}_{\overline{w}}}~.
\end{equation}%
Therefore%
\begin{equation}
dz\wedge d\overline{z}~\sqrt{1-4b^{-2}\ u_{z}\,u_{\overline{z}}}~=dw\wedge d%
\overline{w}~\sqrt{(z_{\overline{w}}~\overline{z}_{w}-z_{w}~\overline{z}_{%
\overline{w}})^{2}-b^{-2}\left( \overline{z}_{w}-\overline{z}_{\overline{w}%
}\right) \left( z_{\overline{w}}-z_{w}\right) }~.  \label{expression}
\end{equation}%
Since $u(z,\overline{z})$ is the sole degree of freedom in the action (\ref%
{actionuz}), we are free to choose the partner coordinate $v(z,\overline{z}%
)$ in the most convenient way. So, we can look for a choice that
simplifies the
result (\ref{expression}). Then, we will choose the complex coordinate $%
w=u(z,\overline{z})+i~v(z,\overline{z})$ in such a way that the vector $%
\partial /\partial w$ gets the form%
\begin{equation*}
2b\ \frac{\partial }{\partial w}=\xi \ \frac{\partial }{\partial z}+\frac{1}{%
\xi }\ \frac{\partial }{\partial \overline{z}}
\end{equation*}%
for some auxiliary function $\xi $. Thus
\begin{equation}
\frac{\partial z}{\partial w}=\frac{\xi }{2b}~,\hskip2cm\frac{\partial z}{%
\partial \overline{w}}=\frac{1}{2b\,\overline{\xi }}~.  \label{condition}
\end{equation}%
Our choice for $w(z,\overline{z})$ implies that $(u,v)$ are orthogonal
coordinates. In fact,%
\begin{eqnarray}
\frac{\partial }{\partial u}\cdot \frac{\partial }{\partial v} &=&i\ \left(
\frac{\partial }{\partial w}+\frac{\partial }{\partial \overline{w}}\right)
\cdot \left( \frac{\partial }{\partial w}-\frac{\partial }{\partial
\overline{w}}\right) =i\ \frac{\partial }{\partial w}\cdot \frac{\partial }{%
\partial w}-i\ \frac{\partial }{\partial \overline{w}}\cdot \frac{\partial }{%
\partial \overline{w}}  \notag \\
&=&i\ (g_{ww}-g_{\overline{w}\overline{w}})=2\,i\ g_{z\overline{z}}\left(
\frac{\partial z}{\partial w}\frac{\partial \overline{z}}{\partial w}-\
\frac{\partial z}{\partial \overline{w}}\frac{\partial \overline{z}}{%
\partial \overline{w}}\right) \ =\frac{i\,}{4b^{2}}\left( \frac{\xi }{\xi }-%
\frac{\overline{\xi }}{\overline{\xi }}\right) =0~.
\end{eqnarray}%
The form of $z_{w}$, $z_{\overline{w}}$ given in Eq.~(\ref{condition}) is
replaced in (\ref{expression}) to obtain
\begin{equation}
dz\wedge d\overline{z}~\sqrt{1-4b^{-2}\ u_{z}\,u_{\overline{z}}}~=\frac{%
dw\wedge d\overline{w}}{4b^{2}}~\left( \frac{1}{\,|\xi |^{2}}-2+|\xi
|^{2}\right) =dw\wedge d\overline{w}~\left( \left\vert \frac{\partial z}{%
\partial \overline{w}}\right\vert ^{2}-2+\left\vert \frac{\partial z}{%
\partial w}\right\vert ^{2}\right) ~.
\end{equation}%
This simple action leads to trivial dynamical equations,%
\begin{equation}
\frac{\partial ^{2}z}{\partial w\,\partial \overline{w}}~=~0~,
\label{equation}
\end{equation}%
which says that $z(w,\overline{w})$ is the sum of a holomorphic function of $%
w$ and an anti-holomorphic function. Notice that the Born-Infeld
constant $b$ is absent in Eq.~(\ref{equation}). Actually this
equation is satisfied by Maxwellian solutions too. However, the
derivatives of $z(w,\overline{w})$ are connected by the
Eq.~(\ref{condition}), which links the holomorphic and the
anti-holomorphic sectors; this is the only role left to $b$.
Therefore the general solution is
\begin{equation}
z~=~f(w)+\frac{g(\overline{w})}{4\,b^{2}}  \label{1}
\end{equation}%
(the factor $4\,b^{2}$ has been chosen for convenience), where functions $f$
and $g$ are linked by the condition
\begin{equation}
\overline{f^{\prime }(w)}\ g^{\prime }(\overline{w})=1~.  \label{2}
\end{equation}%
Ideally, the potential $u(z,\overline{z})=\text{Re}[w(z,\overline{z})]$
would be obtained by inverting the result (\ref{1}). However, we can hardly
solve $w(z,\overline{z})$ from Eq.~(\ref{1}), except for trivial choices for
$f$ that could be deprived of physical interest.

\section{Perturbative expansion}

\label{sec3} The inversion of Eq.~(\ref{1}) can be approached in a
perturbative way. For $b^{2}$ going to infinity we recover the Maxwellian
result (the complex potential is an analytic function of $z$). Then we can
start from a known Maxwellian solution $z=f(w)$ and solve the equations (\ref%
{1}) and (\ref{2}) at the order $b^{-2}$; thus we will find the Born-Infeld
corrections at the lowest order. At the level of Eq.~(\ref{BIz}), this
approach is equivalent to replace the Maxwellian potential in the three
terms containing the factor $b^{-2}$.

\bigskip So we will now focus on getting a differential equation for $w(z,%
\overline{z})$, at the first order in $b^{-2}$, that can be easily
integrable. \ By differentiating Eq.~(\ref{1}) with respect to $z$ and $%
\overline{z}$ we get two equations,
\begin{equation}
f^{\prime }(w)\ \frac{\partial w}{\partial z}+\frac{g^{\prime }(\overline{w})%
}{4b^{2}}\ \frac{\partial \overline{w}}{\partial z}=1~,  \label{3}
\end{equation}%
\begin{equation}
f^{\prime }(w)\ \frac{\partial w}{\partial \overline{z}}+\frac{g^{\prime }(%
\overline{w})}{4b^{2}}\ \frac{\partial \overline{w}}{\partial \overline{z}}%
=0~.  \label{4}
\end{equation}%
Let us conjugate the Eq.~(\ref{4}) to solve $\partial \overline{w}/\partial
z $,
\begin{equation}
\frac{\partial \overline{w}}{\partial z}=-\frac{1}{4b^{2}}\ \frac{\overline{%
g^{\prime }(\overline{w})}}{\overline{f^{\prime }(w)}}\ \frac{\partial w}{%
\partial z}~.  \label{5}
\end{equation}%
Therefore, at the first order in $b^{-2}$, only the first term in Eq.~(\ref%
{3}) is relevant. Then it follows that
\begin{equation}
f^{\prime }(w)\ \frac{\partial w}{\partial z}\simeq 1~.  \label{6}
\end{equation}%
But Eq.~(\ref{1}) means that $f^{\prime }(w)=\partial z/\partial w$.
Therefore, Eq.~(\ref{6}) becomes
\begin{equation}
\frac{\partial z}{\partial w}\ \frac{\partial w}{\partial z}\simeq 1~.
\label{7}
\end{equation}%
Eq.~(\ref{7}) is not trivial, since the potential $w$ is a function of $z$
and $\overline{z}$. It is only trivial in Maxwell's theory, where the
potential is analytic in $z$. Eq.~(\ref{7}) means that such a Maxwellian
property is also guaranteed in Born-Infeld theory at order $b^{-2}$. In Eq.~(%
\ref{6}) we can solve $f^{\prime }(w)$ in terms of $\partial w/\partial z$.
Thus, Eq.~(\ref{2}) gives $g^{\prime }(\overline{w})$ in terms of $\partial
\overline{w}/\partial \overline{z}$,
\begin{equation}
g^{\prime }(\overline{w})~=~\frac{1}{\overline{f^{\prime }(w)}}~\simeq ~
\frac{\partial \overline{w}}{\partial \overline{z}}~.  \label{8}
\end{equation}
Results~(\ref{6}) and~(\ref{8}) are used in Eq.~(\ref{4}) for
getting the differential equation
\begin{equation}
\frac{\partial w}{\partial \overline{z}}+\frac{1}{4b^{2}}\ \frac{\partial w}{%
\partial z}\ \left( \frac{\partial \overline{w}}{\partial \overline{z}}%
\right) ^{2}~=~\mathcal{O}(b^{-4})~.  \label{9}
\end{equation}

\section{Periodic solutions}

\label{sec4} In Eq.~(\ref{9}), let us try the solution
\begin{equation}
w(z,\overline{z})=D\ \cos [k\,z+\Psi (\overline{z})]~,\hspace{0.5in}k\in
\mathbb{R} \ ,  \label{10}
\end{equation}
which is associated with the Maxwellian \textit{seed} $w(z)=D\ \cos (k\,z)$
describing a periodic configuration. In the second term of Eq.~(\ref{9}), we
replace $w$ with its Maxwellian value, without error at the considered
order. Then, one obtains
\begin{equation}
\Psi ^{\prime }(\overline{z})+\ \frac{k^{3}\overline{D}^{2}}{4\,b^{2}}\ \sin
^{2}(k\,\overline{z})~{=~0~.}  \label{11}
\end{equation}%
Thus the complex potential is%
\begin{equation}
w(z,\overline{z})=D\ \cos \left[ k\,z-\frac{k^{3}\overline{D}^{2}}{8\,b^{2}}%
\ \overline{z}+\frac{k^{2}\overline{D}^{2}}{16\,b^{2}}\ \sin [2k\,\overline{z%
}]\right] +\mathcal{O}(b^{-4})~,~  \label{12}
\end{equation}%
which can be simplified by just keeping terms of order $b^{-2}$ in the
development of the cosine function,%
\begin{equation}
w(z,\overline{z})\simeq D\ \cos \left[ k\,z-\frac{k^{3}\overline{D}^{2}}{%
8\,b^{2}}\ \overline{z}\right] -\frac{k^{2}D\overline{D}^{2}}{16\,b^{2}}\
\sin [2k\,\overline{z}]\ \sin \left[ k\,z-\frac{k^{3}\overline{D}^{2}}{%
8\,b^{2}}\ \overline{z}\right] ~.  \label{13}
\end{equation}%
This solution becomes Maxwellian in the limit $k^{2}D^{2}<<b^{2}$. In spite
of appearances, the approximated solution (\ref{13}) is not entirely
satisfactory because the terms of order $b^{-4}$ in the r.h.s. of Eq.~(\ref%
{9}) will include an unbounded term that is linear in $z$. Thus the obtained
solution is valid only for small values of $|z|$. This is a common problem
in perturbative expansions of solutions to differential equations. In
Born-Infeld theory this problem can be solved by resorting to a symmetry
displayed by the equation at the lowest order in $b^{-2}$. Since $\partial
w/\partial z$ only appears in Eq.~(\ref{9}) at the order $b^{-2}$, then the
replacement%
\begin{equation}
\overline{z}~\longrightarrow \overline{z}+\frac{\alpha }{b^{2}}\ z
\label{14}
\end{equation}%
would not have consequences for the solution at the considered order.
Moreover, by choosing $\alpha =-D^{2}k^{2}/8$ we obtain the potential
\begin{equation}
w(z,\overline{z})~\simeq ~D\ \cos \left[ \left( 1+\frac{k^{4}D^{2}\overline{D%
}^{2}}{64\ b^{4}}\right) \,k\,z-\frac{k^{3}\overline{D}^{2}}{8\,b^{2}}\
\overline{z}\right] -\frac{k^{2}D\,\overline{D}^{2}}{16\,b^{2}}\ \sin \left[
2\,k\,\left( \overline{z}-\frac{k^{2}D^{2}}{8\ b^{2}}\ z\right) \right] \
\sin \left[ k\,z-\frac{k^{3}\overline{D}^{2}}{8\,b^{2}}\ \overline{z}\right]
\ ,~  \label{15}
\end{equation}%
which satisfies
\begin{equation}
\frac{\partial w}{\partial \overline{z}}+\frac{1}{4b^{2}}\,\frac{\partial w}{%
\partial z}\left( \frac{\partial \overline{w}}{\partial \overline{z}}\right)
^{2}=~\frac{k^{5}D\overline{D}^{2}}{128\ b^{4}}\,\cos [kz]\,\sin [2k%
\overline{z}]\ \left( \overline{D}^{2}\cos [2k\overline{z}]-4D^{2}\sin
^{2}[kz]\right) +\mathcal{O}(b^{-6}).  \label{16}
\end{equation}%
The terms of order $b^{-4}$ in the r.h.s. are bounded;\ so the solution (\ref%
{15}) is satisfactory. Therefore, even if we are looking for
solutions at the lowest order in $b^{-2}$, we have to consider the
behavior at the next order to avoid unbounded contributions in the
differential equation. The real part of potential (\ref{15}) is then
a proper solution to the equation (\ref{BI}) at the lowest order in
$b^{-2}$.

\section{The Born-Infeld wave equation}

\label{sec5} Equation~(\ref{BI}) becomes a non-linear wave equation by
replacing $y$ with $-it$. Because of this reason, the periodic solution (\ref%
{15}) can help to find solutions to the BI wave equation%
\begin{equation}
(1\,+\,b^{-2}\,u_{,t}^{2})\ u_{,xx}-2\,b^{-2}\,u_{,t}\ u_{,x}\
u_{,tx}-(1\,-\,b^{-2}\,u_{,x}^{2})\ u_{,tt}\ =\ 0\ .  \label{17}
\end{equation}%
Notice that the real part of the Maxwellian seed is
\begin{equation}
\text{Re}[D\ \cos (k\,z)]=D_{1}\cos (kx)\cos (iky)-iD_{2}\sin (kx)\sin
(iky)\ .  \label{18}
\end{equation}%
If $D_{2}=0$, then the replacement $y\longrightarrow -it$ leads to a
stationary wave of amplitude $D_{1}$. But in general the result of replacing
$y\longrightarrow -it$ will not be real. However, by replacing $%
D_{1}=a_{1}+a_{2}$ and $D_{2}=i(a_{2}-a_{1})$ we obtain two waves of
amplitudes $a_{1}$, $a_{2}$ that propagate towards increasing and decreasing
values of $x$ respectively. We will now start from the potential (\ref{15})
and follow this recipe to obtain the corresponding BI solution at the order $%
b^{-2}$. The result is%
\begin{eqnarray}
&&u(t,x)\ =\ \left\{ a_{1}\ \cos \left[ \left( 1-\frac{a_{2}^{2}\,k^{2}}{%
2b^{2}}\right) \,kx-\left( 1+\frac{a_{2}^{2}\,k^{2}}{2b^{2}}\right) \,kt%
\right] +a_{2}\ \cos \left[ \left( 1-\frac{a_{1}^{2}\,k^{2}}{2b^{2}}\right)
\,kx+\left( 1+\frac{a_{1}^{2}\,k^{2}}{2b^{2}}\right) \,kt\right] \right\}
\notag \\
&&  \notag \\
&&\left\{ 1-\frac{a_{1}a_{2}\,k^{2}}{2b^{2}}\sin \left[ \left( 1-\frac{%
a_{2}^{2}\,k^{2}}{2b^{2}}\right) \,kx-\left( 1+\frac{a_{2}^{2}\,k^{2}}{2b^{2}%
}\right) \,kt\right] \ \sin \left[ \left( 1-\frac{a_{1}^{2}\,k^{2}}{2b^{2}}%
\right) \,kx+\left( 1+\frac{a_{1}^{2}\,k^{2}}{2b^{2}}\right) \,kt\right]
\right\} \ ,  \label{19}
\end{eqnarray}%
which satisfies the Born-Infeld wave equation at first order in $b^{-2}$:
\begin{eqnarray}
&&(1\,+\,b^{-2}\,u_{,t}^{2})\ u_{,xx}-2\,b^{-2}\,u_{,t}\ u_{,x}\
u_{,tx}-(1\,-\,b^{-2}\,u_{,x}^{2})\ u_{,tt}\ =\ -\frac{a_{1}a_{2}\ k^{6}}{%
4\,b^{4}}\ (\cos [2kt]-\cos [2kx])  \notag \\
&&  \notag \\
&&\Big\{a_{2}(4\,a_{1}^{2}+a_{2}^{2})\,\cos
[k(t+x)]+a_{1}(4\,a_{2}^{2}+a_{1}^{2})\,\cos [k(t-x)]+a_{1}^{3}\,\cos
[3k(t-x)]+a_{2}^{3}\,\cos [3k(t+x)]  \notag \\
&&  \notag \\
&&+6\,a_{1}a_{2}^{2}\ \left( \cos [k(3t+x)]+\cos [k(t+3x)]\right)
+6\,a_{1}^{2}a_{2}\ \left( \cos [k(3t-x)]\,+\cos [k(t-3x)]\right) \Big\}\ +%
\mathcal{O}(b^{-6})~  \label{20}
\end{eqnarray}%
(notice that a \textit{bounded} function of order $b^{-4}$ is left).

\section{Born-Infeld electromagnetic waves}

\label{sec6}

The Born-Infeld wave equation (\ref{17}) is involved in some
configurations of
BI electrodynamics. For instance, consider the potential%
\begin{equation}
A_{i}\ =\ (0,\ 0,\ u(t,x),\ 0)\ .  \label{21}
\end{equation}%
Then the components of the field tensor are%
\begin{equation}
F_{ty}=E_{y}=\frac{\partial u}{\partial t}\ ,\hspace{0.5in}F_{xy}=-B_{z}=%
\frac{\partial u}{\partial x}\ ,  \label{22}
\end{equation}%
so $P=0$. Therefore the field $\mathcal{F}^{ij}$ is%
\begin{equation}
\mathcal{F}^{ij}\mathcal{\ }=\ \frac{F^{ij}}{\sqrt{1+\frac{2S}{b^{2}}}}%
\mathcal{\ }=\ \frac{F^{ij}}{\sqrt{1+\frac{1}{b^{2}}\left[ \left( \frac{%
\partial u}{\partial x}\right) ^{2}-\left( \frac{\partial u}{\partial t}%
\right) ^{2}\right] }}\ ,  \label{221}
\end{equation}%
and Eq.~(\ref{BIdyn}) becomes the Eq.~(\ref{17}) for the potential
$u(t,x)$. Therefore, the solution (\ref{19}) can be regarded as the
corresponding BI electromagnetic configuration for the Maxwell field
of two equally polarized monochromatic plane waves propagating along
the $x-$axis in opposite directions. Such a configuration could be
produced when a plane wave is perpendicularly incident on a plane
interface. Since a source distribution will appear at the interface,
we remark that the dynamical equation (\ref{BIdyn}) for the coupling
$A_{j}\,j^{j}$ becomes
\begin{equation}
\partial _{i}\mathcal{F}^{ij}\ =\ 4\pi \ j^{j}\ .  \label{310}
\end{equation}

\subsection{Mirror}

If the interface is a mirror, then the electric field must vanish on
the mirror. This boundary condition is achieved for
$a_{1}=-a_{2}\equiv a$ (assuming that the mirror is located at
$x=0$). In fact, the potential becomes
\begin{equation}
u=2\,a\,\sin \left[ \left( 1+\frac{a^{2}\,k^{2}}{2b^{2}}\right)
kt\right] \ \sin \left[ \left( 1-\frac{a^{2}\,k^{2}}{2b^{2}}\right)
kx\right] \ \left\{ 1+\frac{a^{2}k^{2}}{4b^{2}}\left( \cos \left[
\left( 1+\frac{a^{2}\,k^{2}}{2b^{2}}\right) 2kt\right] -\cos \left[
\left( 1-\frac{a^{2}\,k^{2}}{2b^{2}} \right) 2kx\right] \right)
\right\} \ .  \label{23}
\end{equation}
So, we obtain a stationary wave where $E_{y}=\partial u/\partial t$
has nodes at positions $x_{n}$ given in terms of the arbitrary
wavenumber $k$ as follows:
\begin{equation}
\left( 1-\frac{a^{2}k^{2}}{2b^{2}}\right) \ k\,\,x_{n}\ =\ n\,\pi \ .
\label{24}
\end{equation}
While in Maxwell's theory the consecutive nodes are separated by a
half-wavelength, in BI electrodynamics the separation between nodes
is affected by the wave amplitude (see References
\cite{Ferraro3,Ferraro4} for the case of waveguides).

\bigskip

Equation (\ref{23}) also shows that the amplitude of the stationary
wave is modulated by its second harmonic. In other words, at the
order $b^{-2}$ the field contains the fundamental frequency and its
third harmonic. The amplitude of the modulation,
$a^{2}k^{2}/(4b^{2})$, is expected to be very weak. This tiny
non-linear effect could be evidenced by a sensitive measurement of
the electric current in the mirror surface. According to
Eq.~(\ref{310}) a current $j^{y}=g(t){\delta }(x)$ is associated
with the jump of the magnetic component $\mathcal{F}^{xy}$. Then, it
follows that
\begin{equation}
g(t)\ =\ \frac{1}{4\,\pi}\,\Delta \mathcal{F}^{xy}|_{x=0}=\
\frac{a\,k}{2\,\pi}\,\sin \left[ \left(
1+\frac{a^{2}k^{2}}{2b^{2}}\right) kt\right] \left(
1-\frac{7\,a^{2}k^{2}}{2\,b^{2}}+ \frac{5\,a^{2}k^{2}}{2\,b^{2}}
\,\cos \left[ \left( 1+\frac{a^{2}k^{2}}{2b^{2}} \right) 2kt\right]
\right) \ .
\end{equation}%
This current could be detectable in the walls of the resonant cavities of
high power lasers.

\subsection{Interface with a magnetostatic uniform field}

It is well known that BI electromagnetic waves travel at a speed
lower than $c$ (we are using $c=1$) in the presence of a
magnetostatic field $\mathbf{B}_{o}$. For a propagation direction
orthogonal to $\mathbf{B}_{o}$, the propagation velocity is
\cite{Plebanski,Boillat,Salazar,Novello,Aiello}
\begin{equation}
\beta \ =\ \frac{1}{\sqrt{1+\frac{B_{o}^{2}}{b^{2}}}}\ .  \label{25}
\end{equation}
In this sense, the presence of a magnetostatic field
$\mathbf{B}_{o}$ makes the space behave like a refractive medium.
Therefore it could be expected that an interface with an (otherwise)
empty semi-space containing a uniform magnetostatic field
$\mathbf{B}_{o}$\textbf{\ }will produce a transmitted wave together
with a \textit{reflected} wave. Notice that Maxwell's theory does
not produce any reflection at the interface; since the theory is
linear there is no interaction between fields.\ So the Maxwell wave
is not affected by the magnetostatic field; $\mathbf{B}_{o}$ just
implies a jump of the magnetic field at the interface, which is
governed by a constant surface current perpendicular to
$\mathbf{B}_{o}$. These features are not so simple in non-linear
Born-Infeld electrodynamics. In fact, the field is governed by
$dF=d(dA)=0$ and the Eq.~(\ref{310}). The last relates the charges
and currents on the interface to the jumps of the normal electric
and tangential magnetic components of $\mathcal{F}$. The former
implies the continuity of the tangential electric and normal
magnetic components of $F$. However, differing from Maxwell's
theory, $\mathcal{F}$ and $F$ do not coincide;\ the electric
(magnetic) part of $\mathcal{F}$ contains traces of the magnetic
(electric) part of $F$ (see the definition of $\mathcal{F}$ in
Eq.~(\ref{BIdyn})). As we will see, this feature is the key to
produce a reflected wave at the interface.

\bigskip

We will consider a normally incident BI plane wave on an interface
at $x=0$. The region $x>0$ (\textit{right} region) contains a
magnetostatic field $B_{o}\,\widehat{\mathbf{y}}$; so, we propose a
constant surface current $\mathbf{j}=g\,\,\delta
(x)\,\widehat{\mathbf{z}}$ at the interface. Since the field $F$ is
purely tangential at the interface, the solution must satisfy the
continuity of $F_{ty}$, $F_{tz}$ and  $\mathcal{F}_{xy}$; besides
the jump of $\mathcal{F}_{xz}$ is governed by the surface current:
\begin{equation}
\,[\mathcal{F}_{right}^{xz}-\mathcal{F}_{left}^{xz}]_{x=0}\,=\,\ 4\,\pi \ g\
.  \label{jump}
\end{equation}
In the \textit{left} region ($x<0$) the form the field is an
incident plane wave polarized along $y$, corrected by a reflected
wave polarized along $z$:
\begin{equation}
F_{left}=-E(x-t)\ d(x-t)\wedge dy+b^{-2}\ H(x+t)\ d(x+t)\wedge dz\
,\label{leftfield}
\end{equation}%
This means that the incident-reflected solution is not related to
the potential (\ref{19}), since the solution (\ref{19}) describes
equally polarized incident and reflected waves. However, as the
reflected wave in the field (\ref{leftfield}) is already of order
$b^{-2}$, the approximated approach will be rather simple in this
case.

In the right region, the transmitted wave polarized along $y$ is
corrected by a transmitted wave polarized along $z$. Both of them
propagate at the
speed $\beta $:%
\begin{equation}
F_{right}=-B(x-\beta \,t)\ d(x-\beta \,t)\wedge dy+B_{o}\ dx\wedge
dz+b^{-2}\ K(x-\beta \,t)\ d(x-\beta \,t)\wedge dz\ .
\end{equation}%
The equation $dF=0$ is trivially fulfilled in both regions. The continuity
of $F_{ty}$ and $F_{tz}$ implies%
\begin{equation}
E|_{x=0}=\ \beta \ B|_{x=0}\ ,\hspace{0.5in}H|_{x=0}=-\beta \ K|_{x=0}\ .
\label{c1}
\end{equation}%
As will be shown, the corrections $H$ and $K$ cannot vanish, but
they are needed to satisfy the Eq.~(\ref{jump}).

\bigskip

Let us compute the field $\mathcal{F}$ as defined in
Eq.~(\ref{BIdyn}). Notice that $H$ and $K$ contribute to $b^{-2}P$
and $(1+b^{-2}\,2S-b^{-4} \,P)^{-1/2}$ at the order $b^{-4}$; thus
it follows that
\begin{equation}
b^{-2}P_{left}=\mathcal{O}(b^{-4})\
,\hspace{0.5in}(1+b^{-2}\,2S-b^{-4}
\,P)_{left}^{-1/2}=1+\mathcal{O}(b^{-4})\ ,
\end{equation}
\begin{equation}
b^{-2}P_{right}=b^{-2}\beta \ B_{o}\ B(x-\beta
\,t)+\mathcal{O}(b^{-4})\ ,
\hspace{0.5in}(1+b^{-2}\,2S-b^{-4}\,P)_{right}^{-1/2}=\beta
+\mathcal{O} (b^{-4})\ .  \label{311}
\end{equation}%
Then $\mathcal{F}_{left}\ \simeq \ F_{left}$, and\footnote{$\ast
\mathcal{F} \simeq B\ d(x-\beta \,t)\wedge dz-\beta \,B_{o}\
dt\wedge dy-b^{-2}\beta ^{2}\,B^{2}B_{o}\ d(x-\beta \,t)\wedge
dy+b^{-2}\beta \,K\ d(\beta \,x-t)\wedge dy$. The equation $d\ast
\mathcal{F}=0$, which is equivalent to the dynamical equation
(\ref{BIdyn}), is trivially satisfied by any differentiable
functions $B(x-\beta t)$, $K(x-\beta t)$ because $d(\beta \,x-t)$
can be replaced by $d(x-\beta \,t)$ without error at the considered
order.}
\begin{eqnarray}
\mathcal{F}_{right}\  &\simeq &\ \beta \ (F-b^{-2}\ P\ \ast F)_{right}\
\notag \\
&\simeq &\ B\ d(t-\beta \,x)\wedge dy+\beta \,B_{o}\ dx\wedge dz+b^{-2}\beta
^{2}\,B^{2}B_{o}\ d(t-\beta \,x)\wedge dz+b^{-2}\beta \,K\ d(x-\beta
\,t)\wedge dz.  \label{312}
\end{eqnarray}
The continuity of $\mathcal{F}_{xy}$ is guaranteed
by the Eq.~(\ref{c1}). The Eq.~(\ref{jump}) implies%
\begin{equation}
\lbrack \beta \,B_{o}-b^{-2}\beta ^{3}\,B^{2}\, B_{o}+b^{-2}\beta
\,K-b^{-2}\ H]_{x=0}\,=\ 4\,\pi \ g\ .
\end{equation}%
Because of Eq.~(\ref{c1}), one gets%
\begin{equation}
\lbrack \beta \,B_{o}-b^{-2}\beta\,E^{2}\, B_{o}-2\, b^{-2} \,
H]_{x=0}\,=\ 4\,\pi \ g\ .
\end{equation}
Let us introduce a monochromatic incident wave $E=e\,\sin [k(x-t)]$.
Thus
\begin{equation}
\beta \,B_{o}-\frac{1}{2}\ b^{-2}\,\beta\ e^{2}\, B_{o}\,
\left(1-\cos[2kt]\right)-2\, b^{-2}\, [H]_{x=0}\ =\ 4\,\pi \ g\ .
\end{equation}
Then
\begin{equation}
\beta \,B_{o}\ \left(1-\frac{e^{2}}{2\, b^2}\right)\ =\ 4\,\pi \ g\
,
\end{equation}
i.e.,
\begin{equation}
\,B_{o}\ \left(1-\frac{B_o^2+e^{2}}{2\, b^2}\right)\ \simeq\ 4\,\pi
\ g\ ,
\end{equation}
and%
\begin{equation}
\frac{1}{2}\ b^{-2}\,\beta\ e^{2}\, B_{o}\, \cos[2kt]-2\, b^{-2}\,
[H]_{x=0}\ =0\ .
\end{equation}
Therefore the reflected wave is
\begin{equation}
b ^{-2}\,H(t+x)\ = \frac{e^2\, \beta B_{o}}{4\ b^2}\ \cos[2k(x+t)]\
\simeq\ b^{-2}\, \pi\, e^2\, g\ \cos[2k(x+t)]\ .
\end{equation}
According to Eq.~(\ref{c1}), the transmitted wave is given by
\begin{equation}
B(x-\beta\, t)\ =\ \beta^{-1}\, e\,\sin\left[\frac{k}{\beta}\,
(x-\beta\, t)\right]\hskip.5cm\textrm{and}\hskip.5cm b
^{-2}\,K(x-\beta\, t)\ = -\frac{e^2\, B_{o}}{4\ b^2}\
\cos\left[\frac{2\, k}{\beta}\,(x-\beta\, t)\right]\ .
\end{equation}

\section{Conclusions}

\label{sec7} Exact solutions to the Born-Infeld electrostatic
equation (\ref{BI}) (or, equivalently, (\ref{BIz})) are linked to a
pair of holomorphic and anti-holomorphic functions, $f$ and $g$
respectively, subjected to the condition (\ref{2}). Nevertheless the
solutions remain expressed in the implicit way (\ref{1}), where the
coordinates $z$ are given as functions of the complex potential $w$.
This relation can be inverted order by order in powers of $b^{-2}$,
to get the (real) potential as a function of the coordinates. This
procedure is equivalent to solve the equations order by order; in
this case, the equations are satisfied up to terms of higher order
in powers of $b^{-2}$. However such terms could develop secular
behaviors for large values of the coordinates. We have taken
advantage of the symmetry (\ref{14}) to avoid the undesired
behavior. The solutions to the scalar equations (\ref{BI}) and
(\ref{17}) can be applied to obtain simple configurations of
Born-Infeld electrodynamics: those where $\mathbf{E}$ and
$\mathbf{B}$ are orthogonal and depend just on two coordinates (a
third coordinate can always be introduced by means of a Lorentz
boost). With this aim, we considered periodic complex potentials for
getting a configuration of electromagnetic BI waves traveling in
opposite directions (see Reference \cite{Barbashov} for a different
approach to this problem). In the case of the stationary wave
produced by the reflection of an BI electromagnetic wave normally
incident on a mirror, we have shown that the separation between
nodes becomes dependent on the amplitude of the wave. Besides, both
the wave amplitude and the surface current at the mirror are
modulated by the second harmonic. These effects could be detected in
cavities, and have consequences in the phenomenology of
laser-plasmas \cite{Burton}. We have also studied the propagation of
BI electromagnetic waves at the interface with a region containing a
magnetostatic field. It is well known that the magnetostatic field
alters the propagation velocity, like a sort of refractive index;
this property is exploited in some experimental arrays recently
proposed \cite{Perlick}. We have shown that the analogy with a
refractive index also works to produce a reflected wave at the
interface, whose polarization is orthogonal to the one of the
incident wave. Irrespective of the technical difficulties in
carrying out this experiment, we should keep the idea that it is
possible to test BI electrodynamics by searching for optical effects
in (non-necessarily homogeneous) static fields; these effects
include deviation of rays and contributions of harmonic frequencies.

\begin{acknowledgments}
This work was supported by Consejo Nacional de Investigaciones Cient\'{\i}ficas y
T\'{e}cnicas (CONICET) and Universidad de Buenos Aires.
\end{acknowledgments}

\end{document}